\documentclass[twocolumn]{aastex701}

\usepackage{graphicx}
\usepackage{eurosym}
\usepackage{natbib}
\usepackage{booktabs}
\usepackage{amsmath}
\usepackage{graphics,graphicx}
\usepackage{hyperref}
\hypersetup{urlcolor=blue}
\usepackage{cleveref}
\usepackage{eurosym}
\usepackage{booktabs}
\usepackage{wrapfig}
\usepackage{tabularx}

\newcommand{\rsun}{$R_\odot$}
\newcommand\be{\begin{equation}}
\newcommand\ee{\end{equation}}

\def\PP{{\cal P}}

\def\PP{{\cal P}}

\def\la{\mathrel{\mathchoice {\vcenter{\offinterlineskip\halign{\hfil
$\displaystyle##$\hfil\cr<\cr\sim\cr}}}
{\vcenter{\offinterlineskip\halign{\hfil$\textstyle##$\hfil\cr<\cr\sim\cr}}}
{\vcenter{\offinterlineskip\halign{\hfil$\scriptstyle##$\hfil\cr<\cr\sim\cr}}}
{\vcenter{\offinterlineskip\halign{\hfil$\scriptscriptstyle##$\hfil\cr<\cr\sim\cr}}}}}

\begin{document}

\title{Latitude-Dependent Time Variations of the Solar Tachocline}

\shorttitle{Latitude-dependent solar tachocline variations}

\author[orcid=0000-0002-6163-3472, sname=Basu, gname=Sarbani]{Sarbani Basu}
\affiliation{Department of Astronomy, Yale University, PO Box 208101, New Haven, CT 06520-8101, USA}
\email[show]{sarbani.basu@yale.edu}
\correspondingauthor{Sarbani Basu}

\author[orcid=0000-0003-1531-1541, sname=Korzennik, gname=Sylvain]{Sylvain G. Korzennik}
\affiliation{Center for Astrophysics $|$ Harvard \& Smithsonian, Cambridge, MA 02138, USA}
\email[]{skorzennik@cfa.harvard.edu}

\author[orcid=0000-0002-4995-6180, sname=Tripathy, gname=Sushanta]{Sushanta C. Tripathy}
\affiliation{National Solar Observatory, 3665 Discovery Dr., Boulder, CO 80303, USA}
\email[]{stripathy@nso.edu}

\begin{abstract}

We have examined how the characteristics of the tachocline --- i.e., the change in rotation rate $\delta\Omega$, or the ``jump'', the position of the midpoint of the tachocline, $r_d$, and the width of the tachocline, $w_d$, --- change as a function of time at different latitudes  using 30 years of helioseismic data obtained by the GONG network. We find a  statistically significant change in the jump,  however, these changes do not have a simple  correlation with solar activity. The dependence is different for solar Cycles~23 and 24, and for Cycle~25, it is more similar to that of Cycle~24.  {While our measured changes of the tachocline's width with time are marginally statistically significant, {the cross correlation is statistically significant and   implies that the width is larger when the solar activity is smaller}, suggesting that magnetic fields play a role in confining the tachocline. The position of the tachocline shows a 
significant secular change at low latitudes ($\la 50^\circ$).} At these latitudes, the tachocline has been moving steadily closer to the base of the convection zone. This is consistent with other measurements that have shown that the overall complexity of solar activity has been decreasing over the last few decades. It leads us to speculate that strong magnetic fields tend to push the tachocline deeper into the radiative zone.

\end{abstract}


\section{Introduction} 
\label{sec:intro}

Inversions of helioseismic data have shown that the convection zone (CZ) of the Sun rotates differentially, while the radiative interior rotates like a solid body \citep[e.g.,][etc.]{Schou_1998}. Connecting the two zones is a thin shear layer, known as the tachocline.

While neither the role nor the origin of the tachocline is well understood, a set of solar dynamo models depends on the strong shear in this region to convert weak poloidal fields into strong toroidal fields \citep[e.g.,][and  references therein]{dikpati1999, chatterjee2004, guerrero2008}. 
If indeed the tachocline is the location of the solar dynamo  where magnetic fields are generated, one would expect solar-cycle related changes in the properties of the tachocline.

\citet{basuantia2019} showed that the change in the rotation rate between the solar CZ and the radiative zone (the ``jump'') is a function of time; however, the temporal dependence is not a simple correlation with magnetic activity. The jump followed different trends during solar Cycle~23 than solar Cycle~24 and, furthermore, had different values at the same level of solar activity. \citet{basukorz} found a similar result by analyzing only the most dominant component of the tachocline that showed a $\cos^2\vartheta$ variation with the colatitude $\vartheta$. They further speculated that there is a longer cycle --- longer than either the 11-year solar cycle or the 22-year
magnetic cycle --- that modulates the tachocline jump.

\citet{basuantia2019} did not detect any changes in the position of the
tachocline or the width of the tachocline because of the magnitude of the
uncertainties caused by large error bars in the
helioseismic data they used. These were splitting coefficients derived
using the respective project reduction pipelines, namely from fitting 108-day time series
of observations by the Global Oscillation Network Group \citep[GONG:][]{gong} and
72-day time series of observations by the Michelson Doppler Imager
\citep[MDI:][]{mdi}, an instrument on board the Solar and Heliospheric Observatory (SOHO) and
later from observations by the Helioseismic and Magnetic Imager \citep[HMI:][]{hmi}, an instrument on board the Solar
Dynamics Observatory (SDO). Subsequent analysis using longer data sets \citep{basukorz} did find
significant changes in the width and position of the tachocline; however, as mentioned earlier, 
that work only examined the most dominant splitting coefficients with respect to latitude, which is, of course, an incomplete characterization of the tachocline.

In this paper, we present results of a more complete, two-dimensional analysis of changes of the properties of the tachocline using approximately 2-year-long time series. \citet{basukorz} found that although data sets obtained with longer time  series (1152, 2304, and 4608 days)  were more precise, they tended to wash out solar-cycle-related variations. We also use data sets derived from shorter time series for examining changes in the jump of the tachocline, since it is easier to determine that property precisely.

The remainder of this paper is organized as follows. We describe our tachocline model, the fitting method, and data used in Section~\ref{sec:data}. We describe our results in Section~\ref{sec:res}, and discuss and summarize our findings in Section~\ref{sec:disc}.

\section{The Tachocline Model and Data used} \label{sec:data}

At all latitudes, we  model the tachocline as a sigmoid following
\citet{basu1997}, \citet{abc},  and \citet{basukorz}: 
\begin{equation}
    \Omega(r)_{\rm tach}=\frac{\delta\Omega}{1+\exp[{-(r_d-r)/w_d}]},
    \label{eq:tach}
\end{equation}
where $\delta\Omega$ is the jump in the rotation rate between the convection
zone and the interior, $r_d$ is the position of the tachocline, defined as the
midpoint of the transition (or discontinuity), and $w_d$ is the width of the
transition layer. In this model, the rotation rate changes from $0.269\,\delta\Omega$ to $0.731\,\delta\Omega$
between $r=r_d-w_d$ and $r=r_d+w_d$. Note that the sign of $\delta\Omega$ tells us
whether the convection-zone rotation rate is higher than the radiative-zone
rate or vice versa. The
magnitude, $|\delta\Omega|$, tell us how large the jump is.  

{
Note that there is another model of the tachocline, that of \citet{agk1996},  that has been used in the literature,  where the model uses an error function as follows:
\begin{equation}
    \Omega_{\rm tach, Erf}(r)=\frac{\delta\Omega}{2}\left\{ 1+ {\rm erf}\left[2\frac{(r-r^{(e)}_d)}{w^{(e)}_d}\right]\right\},
    \label{eq:erf}
\end{equation}
 where $\delta\Omega$ is the change in the rotation rate across the tachocline and $r^{(e)}_d$ is the midpoint of the tachocline. For this model, $w^{(e)}_d$ is the radial extent over which the model varies from 0.08 to 0.92 of $\delta\Omega$. Using this model does not change the results for the position, while the width, $w^{(e)}_d$, however, is different, and for small widths (as we find in this work, see \S~\ref{subsec:width}), $w^{(e)}_d$ is five times $w_d$; \citet{basu1997} showed that the two definitions give similar results for the other tachocline parameters.
}

Since we are
investigating latitude-dependent changes, the quantities $\delta\Omega$, $r_d$
and $w_d$ are modeled as functions of colatitude, $\vartheta$, as follows:
\begin{equation}
    \delta\Omega=\delta\Omega_3P_3(\vartheta)+ \delta\Omega_5P_5(\vartheta),
    \label{eq:omega}
\end{equation}
\begin{equation}
    r_d=r_{d_1}+r_{d_3}P_3(\vartheta),
    \label{eq:rd}
\end{equation}
and
\begin{equation}
    w_d=w_{d_1}+w_{d_3}P_3(\vartheta),
    \label{eq:w_d}
\end{equation}
where the functions $P_3$ and $P_5$ are
defined as: 
\begin{equation}
\begin{split}
P_3(\vartheta) &=5\cos^2\vartheta-1,\\
P_5(\vartheta) &=21\cos^4\vartheta-14\cos^2\vartheta+1.
\end{split}
\label{eq:p35}
\end{equation}
{
This latitudinal variation is quite similar to that used by \citet{paulchar}, who used a simple $\cos^2\vartheta$ variation, with $r_d(\vartheta)=r_{d,0}+r_{d,1}\cos^2\vartheta$ and $w_d(\vartheta)=w_{d,0}+w_{d,1}\cos^2\vartheta$. We prefer to keep our dependence explicitly in terms of $P_3(\vartheta)$ since that is the angular dependence of the splitting coefficients used, and hence a more fundamental assumption; this also allowed us to test the effect of higher-order coefficients in a previous work \citep{shape}. Changing the latitudinal dependence from that in Eq.~\ref{eq:rd} to that of \citet{paulchar} does not change the results.
}

The complete model is the same as that used by \citet{basuantia2019}:
 \begin{align}
 \Omega(r, \vartheta)=\begin{cases}
 {\Omega_c+ \Omega_{\rm tach}} \\ \quad\quad\quad\quad\mbox{if } r\le0.7R_\odot\\
\Omega_c +B(r-0.7)
+\Omega_{\rm tach}\\
\quad\quad\quad\quad\mbox{if }  0.7 < r\le0.95R_\odot\\
\Omega_c+0.25B -C(r-0.95)
+\Omega_{\rm tach}\\
\quad\quad\quad\quad\mbox{if } r>0.95R_\odot, 
\end{cases}
\label{eq:2d}
\end{align}
where  $\Omega_c$, $B$ and $C$ are free parameters, and $\delta\Omega_{\rm tach}$
is given by Eq.~\ref{eq:tach}.  

We determine the tachocline parameters using simulated annealing
\citep{anneal1,anneal2} to minimize a least-squares fit between the observed
and computed values of the data. This algorithm uses randomly generated initial
values of the model parameters, and we assume that these parameters have
Gaussian priors with their mean and width determined from existing inversions
of rotational splittings. Since there is a finite chance that the solution
becomes trapped in a local minimum, we make 100 different realizations using
different sequences of randomly selected initial guesses in order to determine
a global $\chi^2$ minimum. We confirm whether the algorithm reached a
global minimum by inspecting the likelihood-weighted distribution of all the
parameters for all iterations. This distribution is single-peaked when a
global $\chi^2$ minimum is reached; otherwise, it will have multiple peaks (or
be flat if the parameter cannot be constrained). We determine uncertainties of
the model parameters using the traditional bootstrapping method of simulating
many realizations of the observations and fitting each one of them in the same
manner as the original data, and we use the spread as a measure of uncertainty \citep{Laarhoven+Aarts}. Note that the lower error bars that result from fitting longer time series allow one to fit additional terms in the expansion of $r_d$ with respect to the colatitude \citep[as shown by][]{shape}; however, the 720-day sets do not allow us to do so.

The data used in this work are derived from ground-based solar oscillation 
observations acquired by the GONG instruments. GONG has been observing the Sun without interruption since 1995; hence, these data are ideal for studying time
variations of the solar interior.

Most of the rotational frequency splittings that we use were obtained by an
independent data reduction pipeline, which we refer to as the ``SGK'' pipeline 
\citep{syl1, syl2, syl3, syl4, syl5, sgk2018, sgk2023} after the author of the pipeline. The SGK pipeline
derives mode parameters from time series that are multiples of 72 days. For
this work, we use splittings obtained with 144-day time series and 720-day
time series, with the 720-day data sets having a 360-day overlap. 
While the GONG project does not routinely produce frequencies
with 144-day or 720-day time series, a custom 720-day
non-overlapping set of frequencies and splittings was produced for this work by co-author SCT using the  GONG project pipeline. As mentioned earlier, the data obtained with the shorter time series are used to determine the jump, while the longer time series data are used to determine the other parameters.

The sets  are available in the standard form, i.e., expressed as follows:
\begin{equation}
\nu_{nlm}
= \nu_{nl} + \sum_{j=1}^{j_{\rm max}} c_j (n,l) \, \PP_j^{(l)}(m), 
\label{eq:eq_split}
\end{equation}
where $\nu_{nl}$, the central frequency of a mode of degree $l$ and radial
order $n$, is determined by the spherically symmetric part of the solar structure,
$c_j$ are splitting coefficients, and $\PP_j$ are rescaled Clebsch-Gordon
coefficients \citep[see][]{ritz}. With the oscillation frequencies expressed
in this form, one can show that the odd-order $c_j$ are caused by the solar
rotation and have information about the latitudinal distribution, while the even-order coefficients contain signatures of asphericity
and magnetic fields.  We work with the $c_j$ coefficients as defined by \citet{ritz}; both the SGK and the GONG project pipelines, produce these. 

As with \citet{shape}, we restrict ourselves to using modes with frequencies between 1.5~mHz and 3.5~mHz that have lower turning points between 0.55\rsun\ and 0.85\rsun\ for the
range of degrees that is covered by the data set. Note that the lower turning point is the deepest location a mode penetrates, according to ray theory, and effectively the depth below which the oscillation sensitivity decreases drastically; mode properties are most affected by the structure and dynamics at that depth.
This subset gives good coverage of the tachocline, while keeping the uncertainties low (modes with lower and higher frequencies have larger uncertainties). This choice of modes also removes the need to properly account for the near-surface shear layer in the tachocline model, although, as Eq.~\ref{eq:2d} shows, we do model that layer, albeit crudely.  We opted to use the 10.7 cm radio flux as a measure of solar activity \citep{tapping2, tapping}. 

\section{Results}
\label{sec:res}

\subsection{The jump, $\delta\Omega$, across the tachocline}
\label{subsec:jump}

\begin{figure}
    \centering
    \includegraphics[width=3.25 true in]{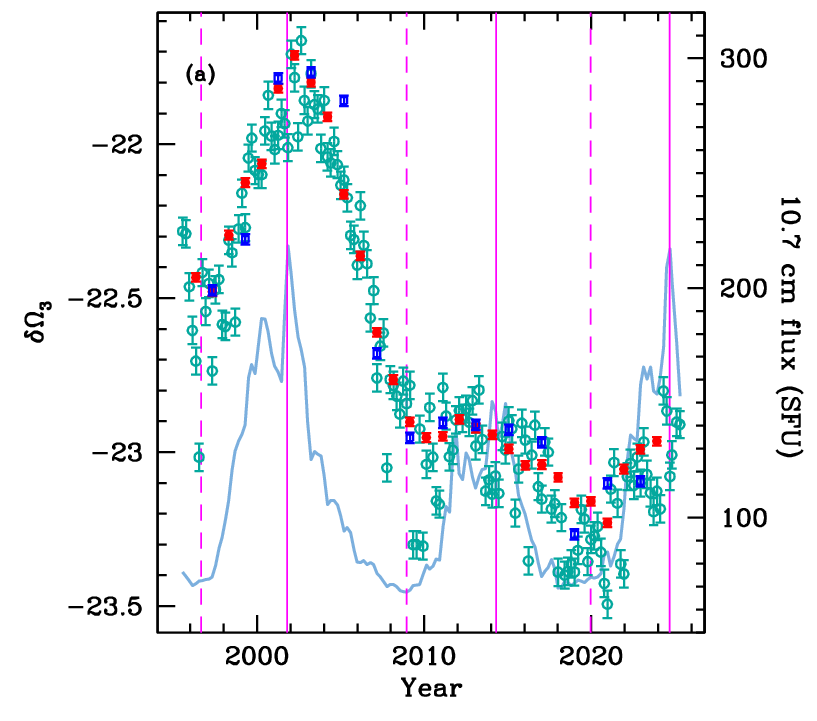}\\
    \includegraphics[width=3.25 true in]{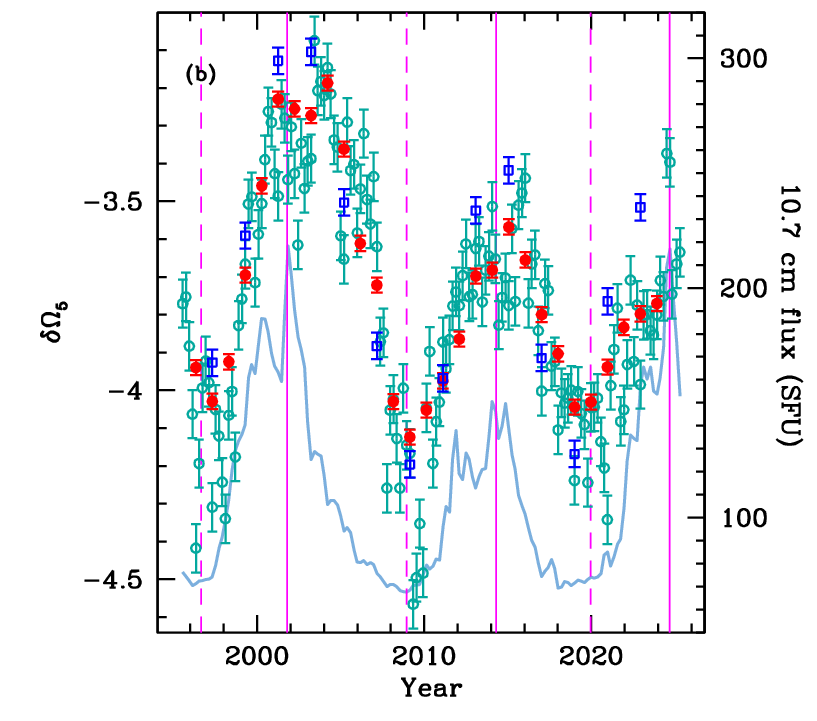}
    \caption{The $\delta\Omega_3$ and $\delta\Omega_5$ components of the change in the
      rotation rate across the tachocline plotted as a function of time, in
      panels labeled (a) and (b), respectively. The greenish-blue points are
      results for 144-day data sets using the SGK pipeline, the red points
      are results for 720-day data sets also using the SGK pipeline, while the
      blue points correspond to results for 720-day data sets using the GONG
      project pipeline.  Values are plotted at the midpoint of the respective
      time series.
      The light blue line shows the 10.7~cm radio flux averaged over the 144 days
      corresponding to each data set. The radio flux magnitude can be read
      from the axes on the right-hand side of the panels. The solid and dashed
      vertical lines are epochs of solar maxima and minima, respectively.  
 }
    \label{fig:j1j2}
\end{figure}
\begin{figure}
    \centering
    \includegraphics[width=3.25 true in]{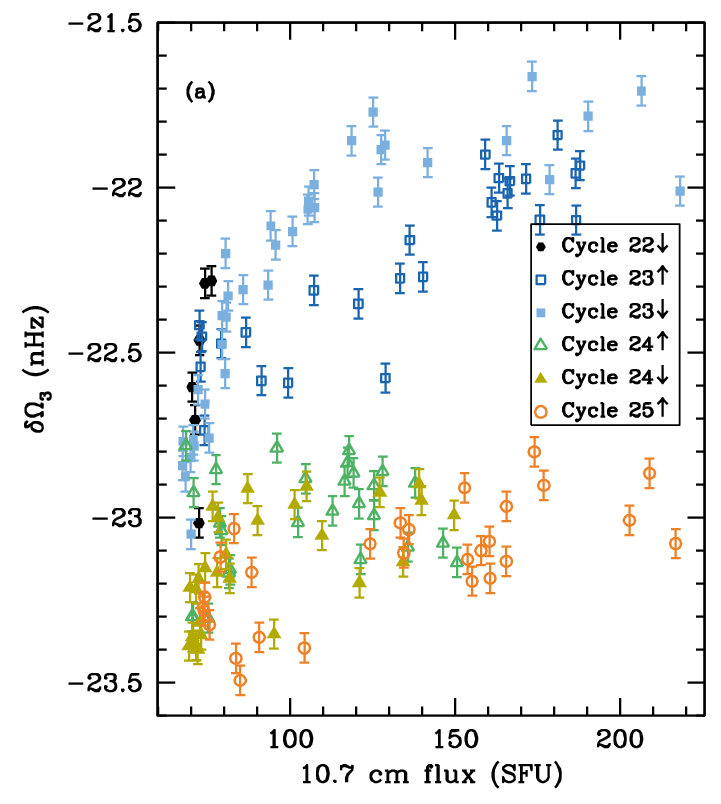}\\
    \includegraphics[width=3.25 true in]{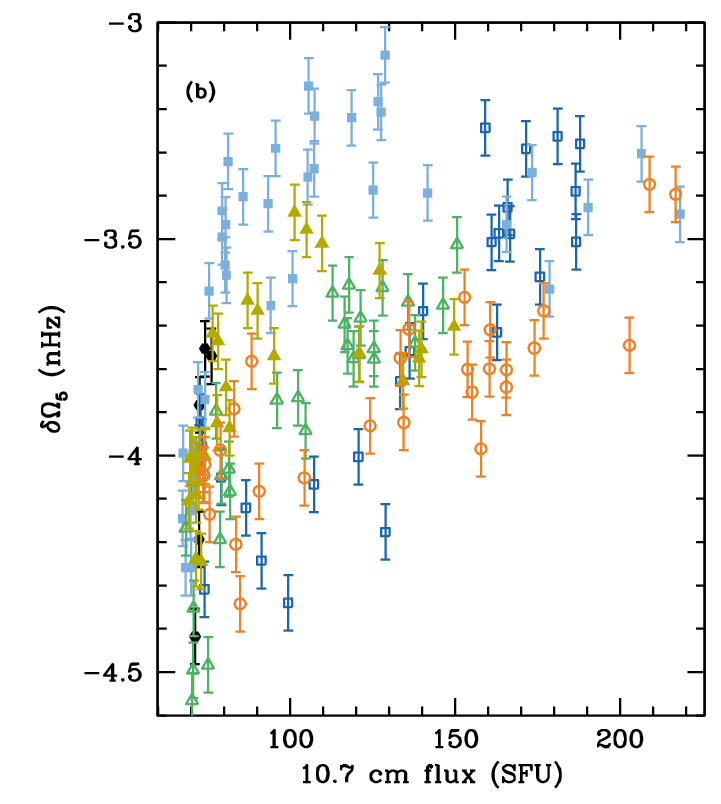}
    \caption{The $\delta\Omega_3$ and $\delta\Omega_5$ components of the change in the
      rotation rate across the tachocline plotted as a function of the 10.7 cm
      radio flux in panels labeled (a) and (b), respectively. Only results
      using the 144-day data sets are shown. The points are color-coded by the phase of
      the solar cycle, as described by the legend in panel (a), where
      the upward and downward arrows correspond respectively to the ascending and
      descending phases of the cycle.}
    \label{fig:j1j2rf}
\end{figure}

We show the two components of the jump in rotation rate across the tachocline, 
$\delta\Omega_3$ and $\delta\Omega_5$,
as a function of time in Fig.~\ref{fig:j1j2}. Note how their variations with time measured
with either the 144-day or the 720-day sets agree with
each other.
Variations measured using data sets resulting from the SGK and the GONG
project pipelines show the same time dependence, though there are small
differences in the magnitude of the jump. In general, the GONG project
pipeline's data sets produce slightly higher values of $\delta\Omega_5$ at the
rising phases of activity and slightly smaller values at declining phases; the
differences in estimates of $\delta\Omega_3$ are more random, though.

What should be noted is that the component $\Omega_3$, which is essentially
determined by the splitting coefficient $c_3$, does not have a simple
solar-cycle dependence. This is in line with what was found by
\citet{basukorz}, which is not surprising, given that they had used
$c_3$ in that work, and also that the rotational frequency splittings are expanded using orthogonal polynomials. 

One can see that the jump, given by
the absolute value of the $\delta\Omega_3$ component, started increasing in the falling phase
of Cycle~23, had a somewhat flat time dependence during the rising phase of
Cycle~24, increased again in the falling phase of that cycle, and started
decreasing in the rising phase of Cycle~25. The second component of the jump,
i.e., $\delta\Omega_5$, on the other hand, rises and falls with solar activity. The
two, however, do not change in lockstep, and there is a time lag, as well as
cycle-to-cycle differences. For instance, $\delta\Omega_5$ kept decreasing
well after the minimum between Cycles~22 and 23, even after Cycle~23 had
started rising, while the minimum in $\delta\Omega_5$ between solar Cycles~23 and 24
occurred at the activity minimum, as it did for the Cycle~24--25 minimum. 

\begin{table}[htb]
    \centering
    \caption{f{The slope and $p$-values obtained for a linear regression of $\delta\Omega$ with respect to the 10.7 cm radio flux. The slope is in units of nHz (SFU)$^{-1}$.}}
    \begin{tabular}{lcccc}
    \hline
    Cycle \& &\multicolumn{2}{c}{$\delta\Omega_3$}& \multicolumn{2}{c}{$\delta\Omega_5$}\\
    Phase& Slope& p-value & Slope & p-value \\
    \hline
      Cycle 23\\
     Ascending & 5.62E$-3$ &  $<$ 1E$-4$ & 7.31E$-3$ & $<$ 1E$-4$ \\
     Descending &  7.33E$-3$ & $<$ 1E$-4$ & 3.99E$-3$ & $<$ 1E$-4$\\
     Cycle 24\\
     Ascending &  1.12E$-3$ & 2.24E$-1$ & 1.00E$-2$ & $<$ 1.3E$-3$\\
     Descending &  4.12E$-3$  & 6.48E$-4$& 5.16E$-3$ & $<$ 7.0E$-4$\\
     Cycle 25\\
     Ascending & 2.64E$-3$ & $<$ 1E$-4$ & 3.90E$-3$ & $<$ 1E$-4$ \\
     \hline     
    \end{tabular}
    \label{tab:regr}
\end{table}

\begin{figure}
    \centering
    \includegraphics[width=3.35 true in]{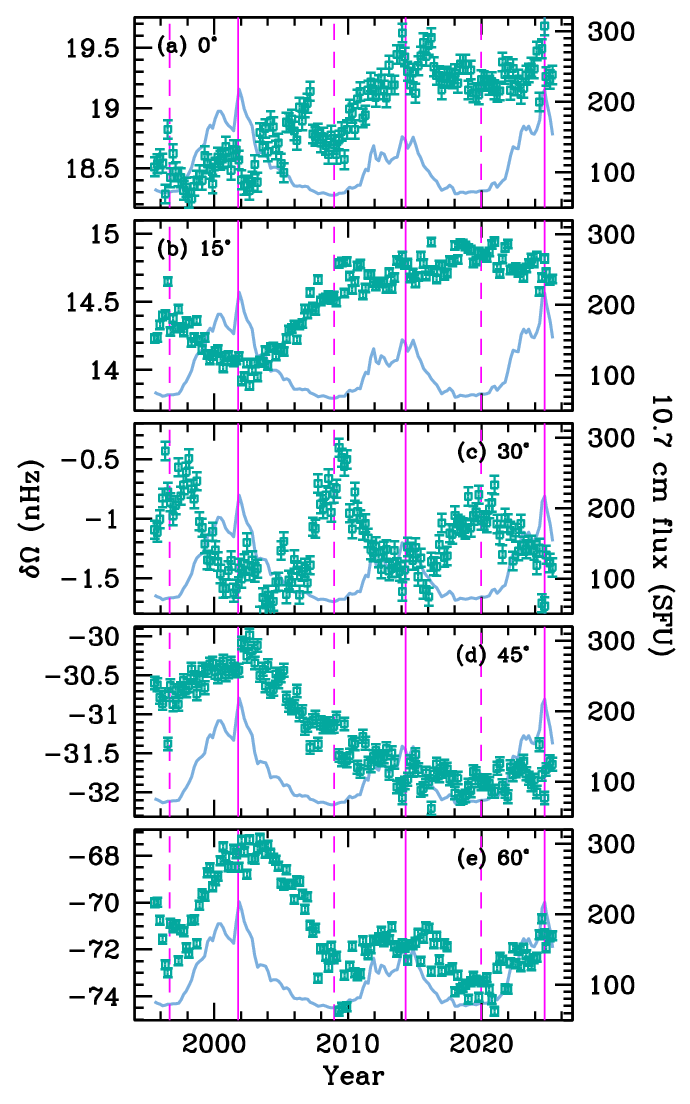}
    \caption{The change in the rotation rate across the tachocline at
      different latitudes plotted as a function of time. Panels (a)--(e) show
      the results at latitudes 0, 15, 30, 45, and $60^\circ$. Only 144-day
      results are shown. The light blue line shows the 10.7~cm radio flux averaged over 144
      days; the magnitude of the radio flux can be read from the axes on the
      right-hand side of the panels. The solid and dashed vertical lines are
      epochs of solar maxima and minima, respectively.}
    \label{fig:j_lat_time}
\end{figure}
\begin{figure}
    \centering
    \includegraphics[width=3.35 true in]{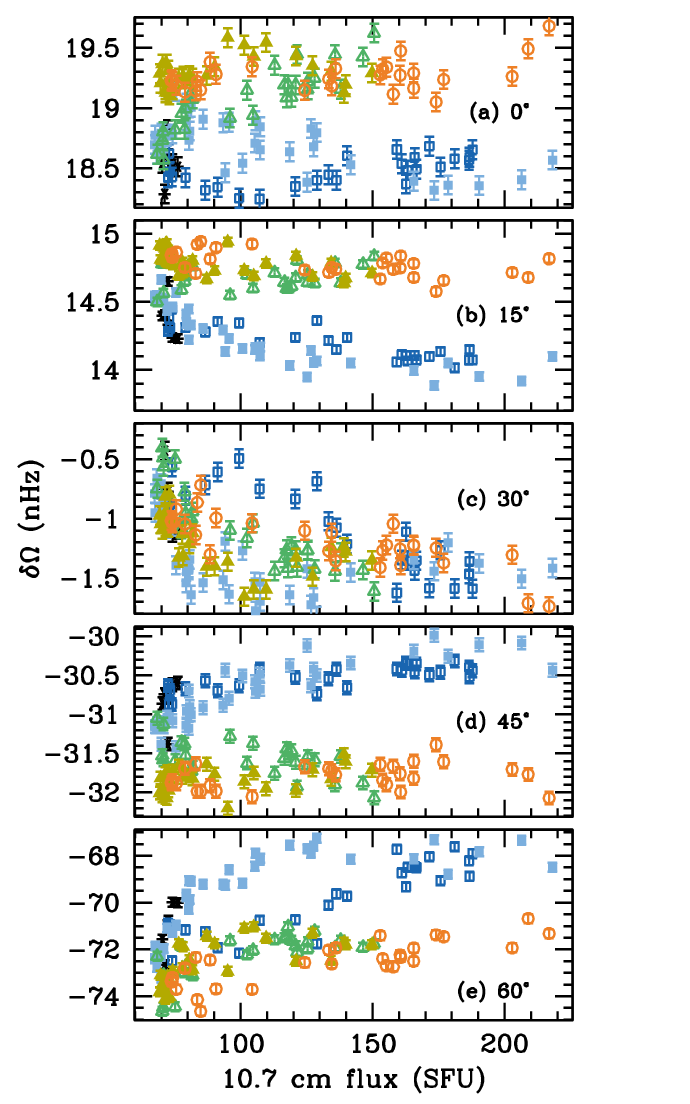}
    \caption{The change in the rotation rate across the tachocline at different latitudes plotted as a function of the 10.7 cm radio flux. Panels (a)--(e) show the results at latitudes 0, 15, 30, 45, and      $60^\circ$. Only 144-day results are shown.  The points are color-coded      by the phase of the solar cycle using the same convention as in      Fig.~\ref{fig:j1j2rf}, i.e., black points for the descending phase of      Cycle~22, dark blue empty squares for ascending Cycle~23, light blue      filled squares for descending Cycle~23, dark green empty triangles for      ascending Cycle~24, light green filled triangles for descending      Cycle~24, and orange points for ascending Cycle~25.}.
    \label{fig:j_lat_rf}
\end{figure}

Both components of the jump show a hysteresis-like behavior with respect to solar
activity, as shown in Fig.~\ref{fig:j1j2rf}. There are two notable
points that we can draw from that figure: first, Cycles 23 and 24 do not show
the same pattern of variation with solar activity levels. This is very clear
for $\delta\Omega_3$, and is also present, though not as clearly, for
$\delta\Omega_5$. Secondly, the pattern for Cycle~25 is similar to that of
Cycle~24, even though it is as strong as Cycle~23. This leads us to speculate
again (see \citealt{basukorz}) that the jump across the tachocline during
Cycle~26 is likely to follow the pattern we see for Cycle~23. In other words,
there might be a four-solar-cycle periodicity in the jump across the
tachocline. Obviously, confirming this hypothesis will require about 10 additional years 
of observations. 

{We have performed a linear regression analysis of $\delta\Omega_3$ and $\delta\Omega_5$ with respect to the 10.7 cm radio flux, and the results are tabulated in Table~\ref{tab:regr}. The lack of correlation of $\delta\Omega_3$ during the rising phase of Cycle~24 is reflected in the corresponding p-value.  }

\begin{figure*}
    \centering
    \includegraphics[width=6.5 true in]{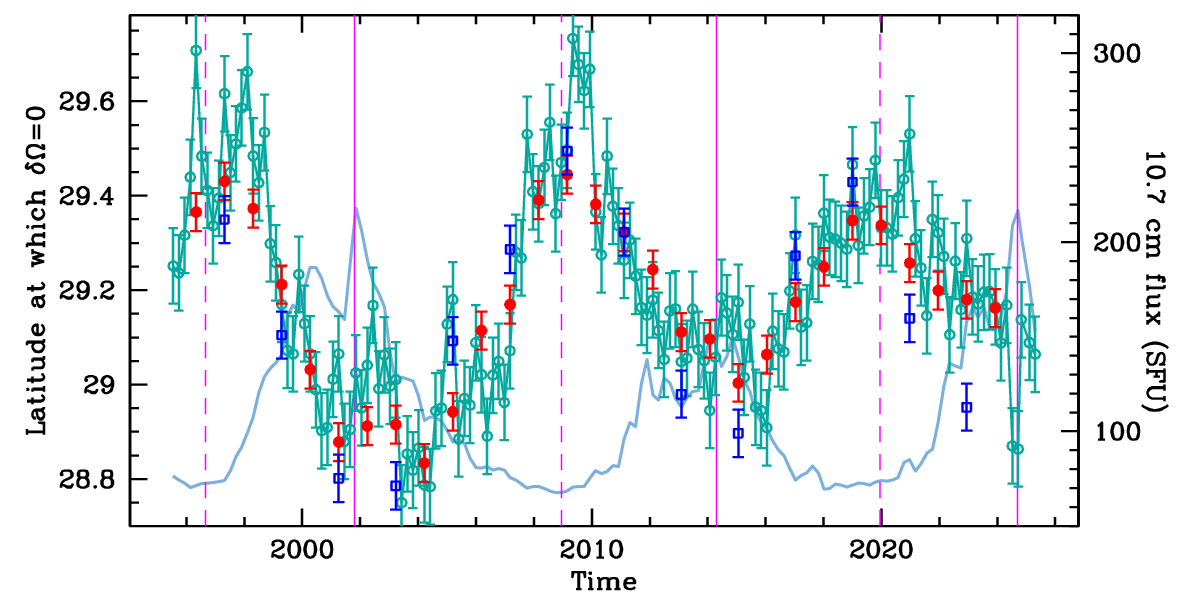}
    \caption{The latitude at which $\delta\Omega=0$, i.e., $\theta_{\rm v}$ plotted as a function of
      time. Greenish-blue points are results for 144-day data sets using the
      SGK pipeline, red points are for 720-day data sets using the SGK
      pipeline, and blue points are for the 720-day data sets obtained with the GONG
      project pipeline. The background light blue line shows the 10.7~cm radio flux averaged over 144 days; the
      magnitude of the radio flux can be read from the axes on the right-hand side
      of the panel. The solid and dashed vertical lines are epochs of solar
      maxima and minima, respectively.}
    \label{fig:zerolat}
\end{figure*}

Although the latitudinal dependence of the jump in the tachocline and changes
thereof is implicit in Figs.~\ref{fig:j1j2} and \ref{fig:j1j2rf}, we
illustrate this dependence explicitly at five latitudes in
Figs.~\ref{fig:j_lat_time} and~\ref{fig:j_lat_rf}.
Fig.~\ref{fig:j_lat_time} shows a clear trend with solar activity only at
$30^\circ$ and $60^\circ$ latitudes. At the latitude of $30^\circ$ the magnitude of the jump
decreases as solar activity decreases, while the opposite is true at the
$60^\circ$ latitude. Of course, we need to keep in mind that the transition as a
function of latitude is quite gentle; the stark change between panels (b),
(c), and (d) is a result of only plotting the jump at a selected few latitudes. The jumps
at different latitudes plotted as a function of activity
(Fig.~\ref{fig:j_lat_rf}) show that at all latitudes the patterns followed by
the jump during different solar cycles are different and that the
patterns seen in Cycle~25 are similar to the ones seen in Cycle~24.

The jump around the latitude of $30^\circ$ is interesting. Indeed, the
conventional view, based on rotation inversions, is that the jump across the
tachocline becomes zero at $30^\circ$. Such inversions, however, often do not
have very good latitudinal resolution \citep[see, e.g.,][]{Schou_1998}, hence,
the actual latitude at which $\delta\Omega$ vanishes, hereafter $\theta_{\rm v}$, is not very well
determined.  Our forward analysis technique  allows us to determine $\theta_{\rm v}$, which we show in Fig.~\ref{fig:zerolat}.  This figure
indicates that $\theta_{\rm v}$ has a clear solar-cycle
dependence; however, this change is not in lockstep with the solar activity
cycle. {New inversion results \citep{sgk_new} indicate that the latitudinal resolution is around $8^\circ$ even when much longer time series data are used; the resolution is expected to be worse for the shorter sets that we are using. Thus, the resolution element in traditional inversions is much larger than the differences that we are seeing, explaining why this intriguing result was not detected earlier.} 

We find that the values of $\theta_{\rm v}$ is generally lower at cycle
maxima and higher at solar minima. Moreover, there is a time lag between the
peaks of activity and the dip in latitude.  Note that the change in $\theta_{\rm v}$
is also affected by the secondary maximum phases --- the two peaks at Cycle~23
maximum appear as two dips; the same is seen
for Cycle~24 maximum, though less clearly. 
{ If we assume that the solar activity levels affect the tachocline, then a cross-correlation analysis indicates that the dips in $\theta_{\rm v}$ lag the peaks in activity by a little more than three years.  }

What is also worth noting is that
there are cycle-to-cycle variances; the position of the dip in $\theta_{\rm v}$ corresponding to
Cycle~23 maximum occurs well after the maximum, but the time lag after
Cycle~24 maximum is much smaller. The situation for the Cycle~25 maximum is
still unclear, although if one looks at the lag between the times of the first
peak in activity and the first dip, one could speculate that Cycle~25 may show
a similar time lag as Cycle~23. Here again, we need a few more years of observations to confirm this.

{To summarize, we find that the jump of the rotation rate across the tachocline, $\delta\Omega$, changes considerably with time, but the change is not a simple function of solar activity. The jump shows a hysteresis like behavior with respect to the 10.7 cm radio flux, with Cycles~23 and 24 following different patterns  while Cycle~25 appears to be mimicking  Cycle~24. We also find that the latitude at which $\delta\Omega$ vanishes, i.e., $\theta_{\rm v}$, shows a significant periodic behavior, being higher at solar minimum and lower at maximum; however, there is a time lag of about 3 years between the minimum of activity and the maximum of $\theta_{\rm v}$. }

\subsection{ The  width, $w_d$, of the tachocline}
\label{subsec:width}

Data obtained with 144-day time series are too noisy to determine the time
variations of the position, i.e., the midpoint of the tachocline, or 
the time dependence of the width of the tachocline; consequently, we only
show results obtained with data from 720-day time series.

\begin{figure}
    \centering
    \includegraphics[width=3.3 true in]{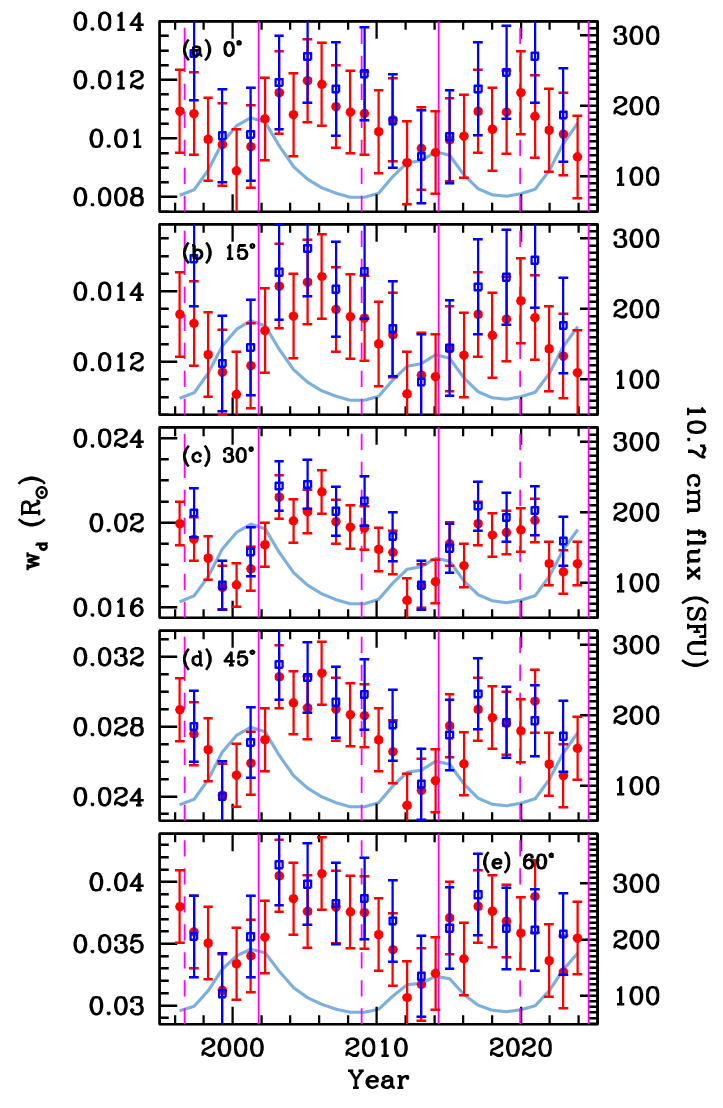}
    \caption{Changes in the width of the tachocline at different latitudes plotted as a function of time.  Panels (a)--(e) show the results at latitudes 0, 15, 30, 45, and $60^\circ$. Only 720-day results are shown.      The red circles are results when using data from the SGK      pipeline, and the blue squares are results when using data from      the GONG project pipeline. The light blue line shows the 10.7 cm radio flux averaged over 720 days corresponding to the SGK data. The magnitude of the radio flux can be read from the axes on the right-hand side of the panel. The solid and dashed vertical lines are  epochs of solar maxima and minima, respectively.}
    \label{fig:width}
\end{figure}

The variation of the width of the tachocline with time is shown in Fig.~\ref{fig:width}. One can see from that figure that the time dependence at all latitudes is more or less the same, but that the width is larger at the higher latitudes. The latter is a well-known result \citep{abc, paulchar, antiabasu2011}, and hence, not surprising.

While the uncertainties in the measured width at each epoch are large, the
variation with time is so smooth that it is conceivable that we have
overestimated these uncertainties. It should be noted that the width of the
tachocline is the parameter that is the most difficult to determine precisely.

We found that the derived estimates of the width depend heavily on the data
uncertainties.  Fitting data with low error bars results in smaller width estimates;
when using data obtained with much longer time series, 
namely 2304-day and 4608-day sets, the resulting value of
the width is quite small \citep[see][]{shape}.

Since all the data sets used in this work have comparable error bars, the
resulting time variation cannot be attributed to their variations. Hence, while the value of the width is likely overestimated, the observed time variation is most likely real.

 {We carried out a number of statistical tests to examine whether the results for the five latitudes shown in Fig.~\ref{fig:width} exhibit significant changes over time. 
We carried out these tests using only the results obtained with SGK data and only for independent sets (i.e., the non-overlapping sets). 
The first was a simple $\chi^2$ test on the residuals resulting from subtracting the mean width at each latitude from the results of that latitude. As expected, this test indicated that the variations were not statistically significant. The next test was a Durbin-Watson test \citep{durbin} on the residuals resulting from performing a linear regression of the data. We first applied it to the time series of the 10.7 cm radio flux, after averaging it over the 720 days corresponding to the SGK data sets to establish a reference  since it is known that the 10.7 cm flux changes with time in a quasi-periodic fashion. The test indicated that changes are significant at the 95\% confidence level. Results of this test on the width variations however, were inconclusive, with the statistic at each latitude lying between the upper and lower confidence limits for a 95\% significance level. The third and final test, a Ljung-Box test \citep{ljung} gave us results that are more interesting. The 10.7 cm radio flux shows significant autocorrelation at time-lags  larger than 2. At latitudes below $45^\circ$, the widths show the same behavior as the 10.7 cm flux, but at higher latitudes, the changes are not significant at the largest lags (11 and above). Therefore, although the uncertainties are large, the tests give us evidence that there is a significant periodic time variation in the tachocline width. A cross-correlation of the 10.7 cm flux and the width at each latitude shows that changes in the 10.7 cm occur about four year before changes in $w_d$, which is why the maximum in $w_d$ appears close to  the minimum in the 10.7 cm flux.}

{To summarize, despite the uncertainties, we find evidence that the width of the tachocline changes in a periodic manner and that the width is larger when activity is smaller. There is a time lag of about four years between the changes in the activity and changes in the width.}

\subsection{ The  position, $r_d$, of the tachocline}
\label{subsec:rad}

\begin{figure}
    \centering
    \includegraphics[width=3.3 true in]{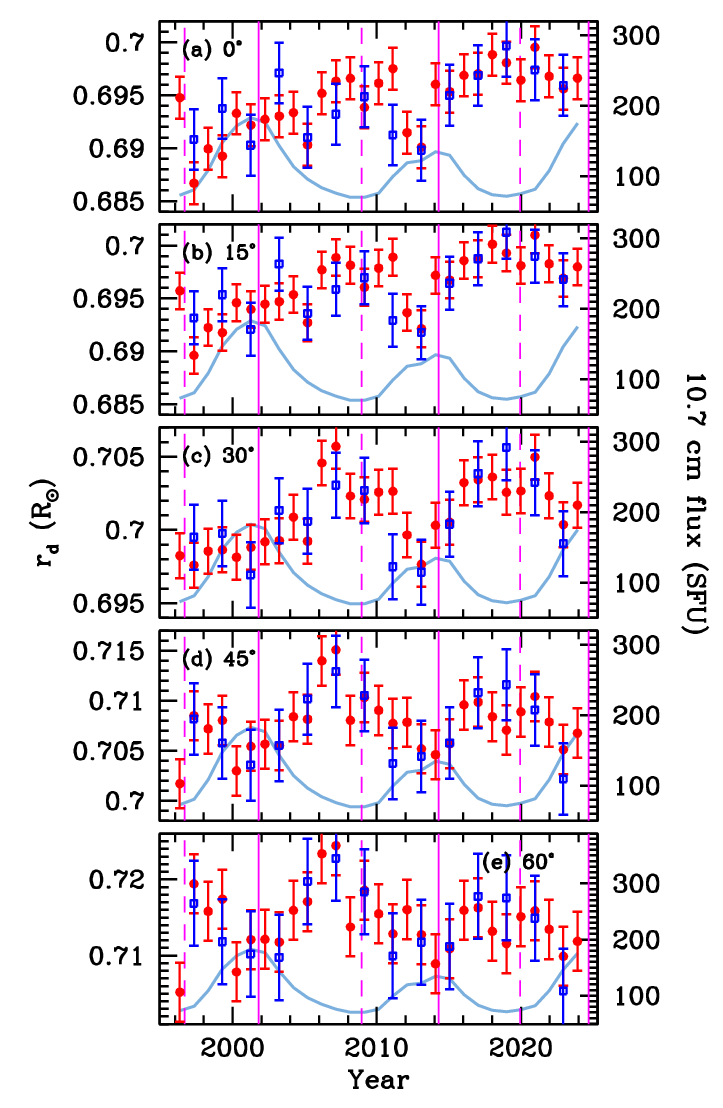}
    \caption{Changes in the position of the tachocline at different latitudes plotted as a function of time. Panels (a)--(e) show the results at latitudes 0, 15, 30, 45, and $60^\circ$. Only 720-day results are
      shown.
      The red circles are results when using data from the SGK
      pipeline, and the blue squares are results when using data from
      the GONG project pipeline.
      The light
      blue line shows the 10.7 cm radio flux averaged over 720 days corresponding to the
      SGK data.
      The magnitude of the radio flux can be read from the axes on the
      right-hand side of the panel. The solid and dashed vertical lines are
      epochs of solar maxima and minima, respectively.}
    \label{fig:position}
\end{figure}

The change in the position of the tachocline is presented in
Fig.~\ref{fig:position} and shows a more random variation, though again the
uncertainties are large; hence, the changes with solar cycle are only
marginally statistically significant.

Despite the large uncertainties, we find that at low latitudes, the
changes are in lockstep with activity levels during Cycle~23 with $r_d$ increasing with
increasing activity, but there appears to be an anti-correlation between
position and activity at higher latitudes. The changeover appears to be around
a latitude of $30^\circ$; around this latitude, the tachocline position was
essentially unchanged during Cycle~23. 
By contrast, for Cycles~24
and 25, the position of the tachocline seems anti-correlated with activity levels, 
with $r_d$ increasing as activity decreases.  There is a
hysteresis-like behavior in the position as well; and just as in the case of the
jump, variations over Cycle~25 follow those of Cycle~24, but the large uncertainties make 
this trend less obvious.

\begin{figure}
    \centering
    \includegraphics[width=3.35 true in]{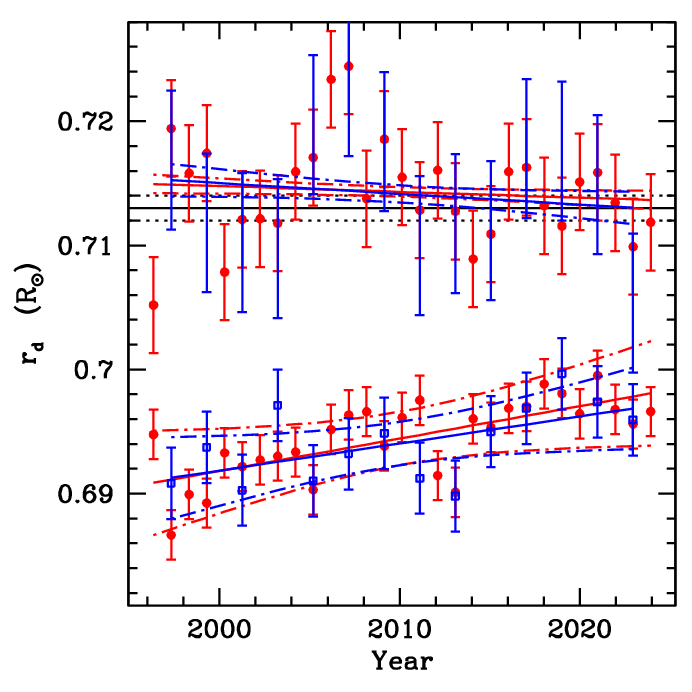}
    \caption{The position of the tachocline at the equator (lower set of points) and at $60^\circ$ latitude (upper set of points) plotted as a function of time. The red points are results with SGK pipeline data, and the blue points are from the GONG project  pipeline. The solid lines of the      corresponding colors show the linear least-squares fit to the points, with the dot-dashed lines indicating the 95\% significance level. The solid black line marks the position of the convection-zone base, with dotted lines showing the $1\sigma$ uncertainty. }
    \label{fig:secular}
\end{figure}

\begin{figure}
    \centering
    \includegraphics[width=3.35 true in]{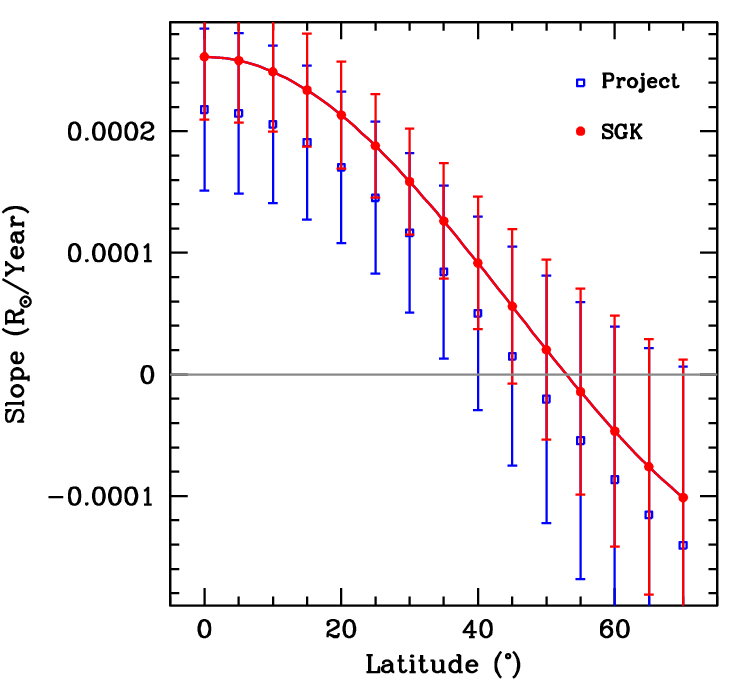}
    \caption{The slope of the temporal variation in the position of the
      tachocline plotted as a function of latitude. The red points are results
      obtained with the SGK pipeline 720-day data, the blue points are
      obtained with the GONG project's pipeline. For reference, a horizontal
      line is shown at zero slope. }
    \label{fig:slope}
\end{figure}

The more interesting variation of the position of the tachocline is the
secular change at low latitudes, as shown in
Fig.~\ref{fig:secular}, while the variation at high latitudes is not statistically
significant.  As shown in the figure, the tachocline has been moving
closer to the base of the convection zone. This secular change can be
fitted with a straight line, and the resulting dependence of this slope 
with latitude is shown in Fig.~\ref{fig:slope}, indicating that it changes sign 
around a latitude of $49^\circ \pm 15^\circ$.

 {As in the case of the width, we applied several statistical tests to examine the significance of the time-variation of $r_d$ at the five latitudes shown in Fig~\ref{fig:position}. Unlike the case of the width, a $\chi^2$ test on the position  show a significant change  at the equator, $15^\circ$ and $30^\circ$ latitudes; the change at $45^\circ$ is at a $1\sigma$ level, while at $60^\circ$ the changes are not significant. The Durbin-Watson test shows that there is no significant autocorrelation in the results, except at $30^\circ$, where the results are inconclusive. Thus the significant change indicated by the $\chi^2$ tests are a result of the secular change in the position.  The Ljung-Box test indicates no significant autocorrelation at the equator or at $60^\circ$, the results at other latitudes are mixed; results at $15^\circ$ and $30^\circ$ show significant autocorrelation (at better than 95\% confidence levels) at lags of 2--4 years, and also for lags greater than 22 years, while at $45^\circ$, the autocorrelations are significant for lags of 6 to 10 years. Cross correlation between the 10.7 cm radio flux and the tachocline position is significant only at $45^\circ$ and $60^\circ$, with a lag of 4 years. Thus, the periodicity in $r_d$ is less significant than that in the tachocline width.
}

{In summary, we find that the position of the tachocline $r_d$ changes with time are significant. The most interesting behavior is, however, the secular change of $r_d$, namely  that the tachocline has moved closer to the convection-zone base over the last 30 years for which we have high quality helioseismic observations. Although we see a hint of periodic behavior that is anticorrelated with solar activity, it is statistically significant only at the higher latitudes. }

\section{Discussions and Conclusions}
\label{sec:disc}

We have examined how the three parameters we use to characterize the tachocline ---
i.e., change in rotation rate $\delta\Omega$, or the jump, the position of the midpoint of
the tachocline, $r_d$, and the width of the tachocline, $w_d$ --- change as a function
of time at different latitudes using 30 years of helioseismic observations acquired by the GONG network. {Earlier work on temporal variation of the tachocline used helioseismic data obtained with 72-day or 108-day time series, but the resulting uncertainties did not enable the detection of time variations in the width and position of the tachocline. To circumvent this, we have used data from longer time series. The jump across the tachocline was determined using rotational splittings obtained with 144-day time series, while the other parameters were obtained with 720-day time series.}

The most statistically significant variation with time is that of the jump. The variation is
different at different latitudes, and does not show a simple correlation with the
level of solar activity. In fact, when plotted as a function of the 10.7 cm radio
flux, Cycles 23 and 24 follow different patterns.  {The rising and falling phases of each cycle, except for the rising phase of Cycle~24, follow a significant linear trend with the 10.7 cm radio flux. } Although Cycle~25 is much
stronger than Cycle~24, the tachocline jump follows the pattern seen  during Cycle~24. This leads us to speculate again (see
\citealt{basukorz}) that the jump across the tachocline during Cycle~26 will follow the pattern seen during Cycle~23, or in other words, there is a four-solar cycle {(i.e., a two Hale-cycle)} period in the jump across the tachocline. Of course, we need about 10 more years of helioseismic data to verify our hypothesis.
{It should be noted that double-Hale cycle variations in the Sun have been reported previously. \citet{efimenko} report a two Hale-cycle period in the power-law index for distributions of equivalent diameters of large groups of sunspots (i.e., sizes of 50–90 Mm.) Earlier, \citet{javaraiah} reported that the equatorial rotation rate of the Sun shows evidence of a double-Hale cycle. }

The latitude at which the jump across the tachocline vanishes, i.e., $\theta_{\rm v}$, the 
location that divides the tachocline into a ``low-$\theta$'' branch where the
convection zone rotates faster than the deep interior and a ``high-$\theta$''
branch where the convection zone rotates slower than the deep interior, changes
with time. This $\theta_{\rm v}$ latitude is lower at solar maximum, leading us
to speculate that such ``sloshing'' of the tachocline might be related to the
magnetic field in the mid-latitude bands. {There is currently no scenario that can explain this behavior. The solar convection zone rotation rate and the tachocline have been modeled by \citet{balbus2009} and \citet{balbusetal2009}, who used the thermal wind equation to deduce the shape of the isorotation surfaces in the bulk of the convection zone and the tachocline. This work was extended further by \citet{balbus}, who showed that the co-latitude  at which the tachocline bifurcated (i.e., where $\delta\Omega=0$, hence at $\theta_{\rm v}$) depended on a parameter that depends on the torque exerted by the tachocline on the interior, in particular at the value at which the viscous torque vanishes. Thus, we may be tempted to say that the changing  $\theta_{\rm v}$ is caused by changing values of the torque; however, as the authors themselves warn, the calculations were done under ideal conditions. Besides, any explanation for the change in $\theta_{\rm v}$ should also explain the lag between $\theta_{\rm v}$ and magnetic activity}. 

The time variation of the width of the tachocline is not statistically
significant {if one relies on simple $\chi^2$ statistics}.  {Other statistical tests reveal the presence of periodicities in the data and a significant cross-correlation signal between  the width and the 10.7 cm radio flux.}  {If we ignore the } large uncertainties, the variation we
see suggests the presence of a thinner tachocline when the Sun is most active. This
agrees with the predictions of \citet{matilsky} who found that Maxwell
stresses arising from a dynamo-generated non-axisymmetric poloidal field
can keep the tachocline from spreading.

\begin{figure}
    \centering
    \includegraphics[width=3.35 true in]{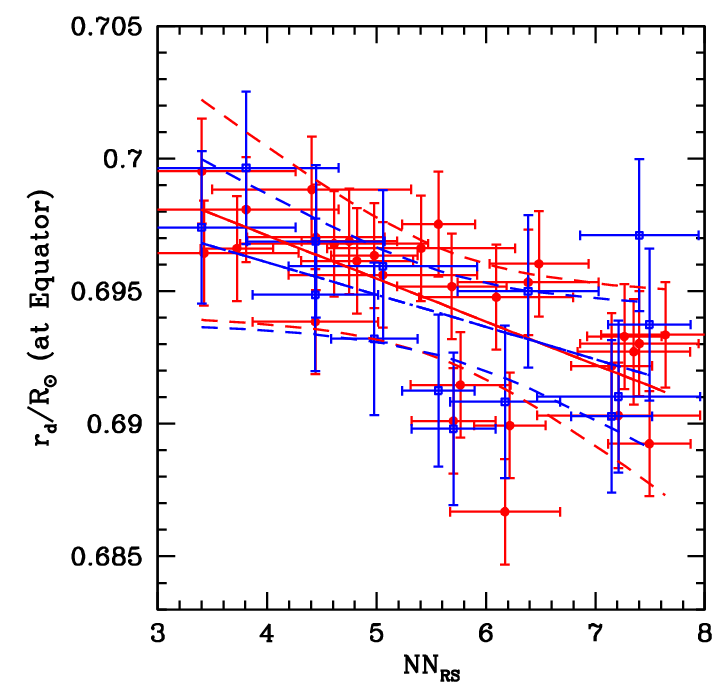}
    \caption{The position, $r_d$, of the tachocline at the equator plotted as a
      function of the annual number of sunspots in a sunspot group. The
       red circles correspond to SGK data, while the blue squares correspond to GONG
       pipeline data. The solid lines show the corresponding linear fit,
      while the dashed lines show 95\% confidence limits. Sunspot data are
     from \citet{alexei}  courtesy of Alexei Pevtsov. }
    \label{fig:ssn}
\end{figure}

Our results on the position of the tachocline, $r_d$, are somewhat more
surprising. Although solar-cycle-related variations are only marginally statistically
significant, it does appear that at low activity levels, the position of the tachocline shifts
to a higher location.
What is surprising, though, is the secular change in $r_d$, where, at low
latitudes, it has been moving to higher radii, and coming closer to
the base of the convection zone. The question that arises naturally, is
whether this is an indication of a weakening of the Sun's magnetic field, or
some property of the magnetic field, over the last few decades. { We believe that there could be a connection with the nature of the solar magnetic field changing over time.} \citet{nago}
found a negative correlation between the numbers of small and large sunspots
during the period of 1998--2011; the number of large sunspots gradually
decreased, while the number of small sunspots steadily increased, indicating a
shift in the properties of the solar magnetic field. \citet{alexei}
found that the annual number of sunspots in a sunspot group shows a steady decline after one removed the solar cycle variation) since about 1990; the authors believe that
the number of sunspots per group may represent active region complexity. As it
happens, the position of the tachocline seems to be correlated with the
average number of sunspots in sunspot groups. We can see this in
Fig.~\ref{fig:ssn} where we plot $r_d$ at the equator as a function of the
average number of sunspots per sunspot group. The relation can be fit with a
straight line, and the correlation is statistically significant.

To conclude, the solar tachocline changes with time. However, the changes are
not a simple function of the solar cycle. There is an indication of a longer
period in the variation of the amplitude of the jump across the tachocline, and there has been a secular change
in the position of the tachocline for the period over which data are
available. The solar-cycle-related changes of the width and position are
barely detectable, even when using data sets obtained with time series that
are almost two years long. 
A more accurate determination of these changes might need other observational approaches and also a better analytical model of the tachocline.

\begin{acknowledgments}
We thank Alexei Pevtsov for providing us with the data on the average number of
sunspots in spot groups.  {We thank the referee for constructive suggestions; we also thank the ApJ statistics editors for suggesting that we perform specific statistical tests.} This work is supported by NASA grant 80NSSC23K0563
to SB and NASA grants 80NSSC22K0516 and NNH18ZDA001N-DRIVE to SGK.  This
work utilizes GONG data obtained by the NSO Integrated Synoptic Program,
managed by the National Solar Observatory, which is operated by the
Association of Universities for Research in Astronomy (AURA), Inc., under a
cooperative agreement with the National Science Foundation and with
contribution from the National Oceanic and Atmospheric Administration
(NOAA). The GONG network of instruments is hosted by the Big Bear Solar
Observatory, High Altitude Observatory, Learmonth Solar Observatory, Udaipur
Solar Observatory,  Instituto de Astrof\'{\i}sica de Canarias, and Cerro Tololo 
Interamerican Observatory.

\end{acknowledgments}

\facilities{GONG, Dominion Radio Astrophysical Observatory}



\bibliography{main}{}
\bibliographystyle{aasjournalv7}

\end{document}